\documentclass[aps,prb,twocolumn,superscriptaddress,showpacs,amsmath,amssymb,footinbib,longbibliography]{revtex4-2}
\usepackage[english]{babel}
\usepackage{graphicx}
\usepackage{dcolumn}
\usepackage{bm}
\usepackage{subfigure}
\usepackage[utf8x]{inputenc}
\usepackage{color}
\usepackage{textcomp}
\usepackage[colorlinks,bookmarks=false,citecolor=blue,linkcolor=red,urlcolor=blue]{hyperref}

\newcommand{\op}[1]{%
    \fontdimen12\textfont3=2pt\fontdimen12\scriptfont3=1.4pt%
    \!\null\mathop{\vphantom{#1}\smash{#1}}\limits_{\sim}\null\!}
\newcommand{\xref}[1]{\protect\ref{#1}}
\newcommand{\figref}[1]{Fig.~\protect\ref{#1}}

\newcommand{\fmref}[1]{(\protect\ref{#1})}

\newcommand{\pp}[2]{\frac{\partial \, {#1}}{\partial \, {#2}}\;}
\newcommand{\Tr}{\mbox{Tr}}

\renewcommand{\eqref}[1]{Eq.~(\protect\ref{#1})}

\newcommand{\vecops}[1]{\op{\vec{s}}_{#1}}
\newcommand{\ops}[1]{\op{s}_{#1}}
\newcommand{\vecopsprod}[2]{\vecops{#1} \cdot \vecops{#2}}

\begin{document}

\title{Magneto- and barocaloric properties of the ferro-antiferromagnetic sawtooth chain}
\author{Nico Reichert}
\author{Henrik Schl{\"u}ter}
\affiliation{Fakult\"at f\"ur Physik, Universit\"at Bielefeld, Postfach 100131, D-33501 Bielefeld, Germany}
\author{Tjark Heitmann}
\affiliation{Fachbereich Mathematik/Informatik/Physik, Universit\"{a}t Osnabr{\"u}ck, Barbarastr. 7, D-49076 Osnabr{\"u}ck, Germany}
\author{Johannes Richter}
\affiliation{Institut f\"{u}r Physik, Otto-von-Guericke-Universit\"{a}t Magdeburg, P.O. Box 4120, D-39016 Magdeburg, Germany}
\affiliation{Max-Planck-Institut f\"{u}r Physik komplexer Systeme, N\"{o}thnitzer Stra\ss e 38, D-01187 Dresden, Germany}
\author{Roman Rausch}
\affiliation{Technische Universität Braunschweig, Institut für Mathematische Physik, Mendelssohnstraße 3, D-38106 Braunschweig, Germany}
\author{J\"urgen Schnack}
\email{jschnack@uni-bielefeld.de}
\affiliation{Fakult\"at f\"ur Physik, Universit\"at Bielefeld, Postfach 100131, D-33501 Bielefeld, Germany}

\begin{abstract}
Materials that are susceptible to pressure and external magnetic fields 
allow the combined use of both for caloric processes. Here we report 
investigations of the ferromagnetic-antiferromagnetic 
sawtooth chain that due to its critical behavior not only allows for both
barocaloric as well as magnetocaloric processes but also features very
large cooling rates in the vicinity of the quantum critical point.

\end{abstract}

\maketitle

\section{Introduction}
\label{sec-1}

Several archetypical frustrated spin systems such as the Heisenberg antiferromagnets
on the kagome, square-kagome, pyrochlore or sawtooth lattice feature a flat energy band
in one-magnon space \cite{Tas:PRL92,Mie:JPA92B,MiT:CMP93}, 
localized one-magnon energy eigenstates, and corresponding multi-magnon energy eigenstates \cite{SSR:EPJB01}. 
These properties result in a macroscopic magnetization jump to saturation \cite{SHS:PRL02}
as well as an increased magnetocaloric effect at the saturation field \cite{ZhH:JSM04},
non-ergodic dynamics \cite{JES:PRB23},
and other frustration effects \cite{MBB:JPCM04}.
For an overview see, e.g., Ref.~\cite{DRM:IJMPB15}.

The saturation field, however, is often not small and thus experimentally hard to reach.
Fortunately, it turns out that for mixed ferromagnetic-antiferromagnetic interaction patterns
models can be set up that feature similar properties at zero field \cite{KDN:PRB14,DmK:JPCM23}.
For the paradigmatic sawtooth (or equivalently delta) chain chemical compounds could be 
synthesized that resemble these properties rather closely \cite{INK:JPSJ05,BML:npjQM18}.
Meanwhile, the class of spin models featuring flat bands has been enlarged by 
moving to special XXZ \cite{CKK:PRL18,CPC:PRB19}
or Dzyaloshinskii-Moriya couplings \cite{ROS:PRB22}.

The critical properties of the ferromagnetic-antiferromagnetic sawtooth chain
with ferromagnetic interaction $J_1 < 0$ 
between nearest neighbor spins
as well as antiferromagnetic 
interaction $J_2 > 0$
between next nearest neighbor spins on even sites, 
compare \figref{delta-structure}, 
can be tuned via the coupling ratio $\alpha = J_2/|J_1|>0$. 
For $0\leq \alpha < \alpha_c = s_a/(2 s_b)$ the ground state is given
by the multiplet of the ferromagnetic state, i.e.,\ with maximal total spin, 
whereas for $\alpha = \alpha_c$ the system features a quantum critical point (QCP)
with macroscopic ground state degeneracy \cite{KDN:PRB14,DmK:PRB15,DSD:EPJB20}. 
For $\alpha > \alpha_c$ the character of the ground state depends on whether the
length of the sawtooth chain (with periodic boundary conditions) is a
multiple of four or a multiple of two but not of four, 
compare \cite{ToK:JMMM04,RPP:SP23}.

Since the density of states varies massively in the vicinity of the 
QCP the system exhibits a huge caloric effect. This will be discussed in  
\xref{sec-3-A}.
Experimentally, the ratio $\alpha$
can realistically be modified by means of pressure, that's why we 
would like to term caloric effects due to a variation of $\alpha$
as barocaloric throughout this article. This dependence was already 
discussed in connection with the molecular ring molecule 
Fe$_{10}$Gd$_{10}$ \cite{BML:npjQM18} and is also observed for
the antiferromagnetic sawtooth chain, see e.g.~\cite{NaT:PLA97}.

Here we want to extend the caloric phase diagram by additionally 
considering variations of the external magnetic field. To this end 
we investigate thermal equilibrium observables ${\mathcal O}(t,h,\alpha)$
as function of reduced temperature $t=T/|J_1|$, 
reduced magnetic field $h=g\mu_B B/|J_1|$,
as well as ``pressure" $\alpha$.
We will demonstrate that the combined use of barocaloric as well 
as magnetocaloric processes can lead to very large cooling rates
in the vicinity of the QCP.

The paper is organized as follows. In Section \ref{sec-2} we introduce 
the model and major observables before we present our results in Sec.~\xref{sec-3}.
The article closes with a discussion in Section~\ref{sec-4}.

\section{Methods}
\label{sec-2}

\subsection{The model}

\begin{figure}[ht!]
\centering
\includegraphics*[width=0.85\columnwidth]{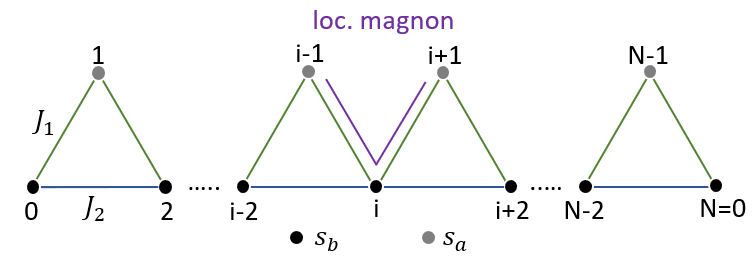}
\caption{Structure of the sawtooth chain with apical spins $s_a$ and basal
spins $s_b$ as well as ferromagnetic interaction $J_1$ and 
antiferromagnetic interaction $J_2$. 
The spins are labeled $0,1,\dots, N-1$.
An independent localized one-magnon state,
that is an eigenstate of the Hamiltonian,
is depicted.}
\label{delta-structure}
\end{figure}

The sawtooth-chain Heisenberg model with periodic boundary 
conditions is given by the Hamiltonian
\begin{align}
    \op{H}^\prime 
    =& J_1 \sum_{i=0}^{N-1} 
    \vecopsprod{i}{i+1} 
    + J_2 \sum_{i=0}^{N/2-1}
    \vecopsprod{2i}{2i + 2}
\label{ham}
\\
&+ g \mu_B B \sum_{i=0}^{N - 1} \ops{i}^z
\nonumber
    \ ,
\end{align}
where $\op{\vec{s}}_i$ denotes the spin vector operator at site $i$, and 
a tilde is used to denote operators in general.
We will consider the case where the exchange coupling 
$J_1 < 0$ is ferromagnetic 
and the exchange coupling $J_2 > 0 $ is antiferromagnetic,
compare also \figref{delta-structure}.

The coupling ratio $\alpha = J_2/|J_1|>0$  
allows to rewrite the Hamiltonian as 
\begin{align}
    \op{H}
    =& 
    - \sum_{i=1}^{N} 
    \vecopsprod{i}{i + 1}  
    +\alpha \sum_{i=0}^{N/2- 1}
    \vecopsprod{2i}{2i+2} 
    \label{hamilton}
    \\
    &+ h \sum_{i=0}^{N - 1} 
    \ops{i}^z
\nonumber
    \ , 
\end{align}
with the dimensionless magnetic field $h =  g \mu_B B/|J_1|$.

In the following we consider $s_a=s_b=s=1/2$. 
Depending on the size $N$ the system is treated with various methods.
For $N\leq 24$ symmetries allow to obtain all energy eigenvalues and eigenvectors 
by means of exact numerical diagonalization of the Hamiltonian 
\cite{HeS:PRB19}.
For $24 < N \lesssim 36$ the finite-temperature Lanczos method (FTLM) allows
to determine low-lying energy eigenvalues as well as thermodynamic observables
with high accuracy \cite{JaP:PRB94,SRS:PRR20}. 
For even larger sawtooth chains with $N>36$ a flavor of the 
density matrix renormalization group theory (DMRG, \cite{Whi:PRL1992,Sch:AP11}) 
which employs SU(2) symmetry is used 
to obtain ground state energies in subspaces of total spin $S$ \cite{RPP:SP23}.
Thermodynamic properties are not investigated for such large systems.

\subsection{Observables}

The observables defined in the following section will be viewed 
as functions of the dimensionless model parameters $\alpha$ 
(related to pressure) 
and $h$ (external field).

Low-temperature caloric figures of merit such as adiabatic temperature change,
isothermal entropy change as well as the cooling rate are massively influenced
by the structure of the low-energy density of states and as a special case 
of that by the degeneracy of the ground state \cite{ZGR:PRL03,SCM:NC14}.
Since the model \fmref{ham} is SU(2)-invariant the total spin of the ground state gives rise 
to a $(2S+1)$-degenerate multiplet at $h=0$ which is Zeeman-split by the magnetic 
field $h$. For instance, this led to a large ground state degeneracy of $(2S+1)=121$
for Fe$_{10}$Gd$_{10}$ since its ground state spin is $S=60$ \cite{BML:npjQM18}.
A ground state degeneracy immediately leads to a residual entropy at zero temperature,
termed $\mathcal{S}_0$ in the following.
Isentropes of smaller entropy run into such points in the $\alpha$-$h$-plane
with decreasing temperature.

Besides the more trivial ground state degeneracy due to the multiplet structure of the 
energy spectrum a massive degeneracy of levels can be observed at the QCP. 
If such a degeneracy
is exponential in $N$, the resulting residual entropy is extensive, 
i.e.\ scales with 
the size of the system $N$. 
This is relevant for macroscopic systems and the thermodynamic 
limit.

Since entropy $\mathcal{S}(t,h,\alpha)$ is a function of 
temperature, field and pressure,
\begin{align}
\mathcal{S}(t,h,\alpha)
&=
- k_B
\Tr
\left[\op{\rho} \log
\left(\op{\rho}
\right)
\right]
\ ,
\label{entropy}
\end{align}
with $\op{\rho}=\op{\rho}(t,h,\alpha)$ being the density operator,
two adiabatic cooling rates can be defined
\begin{align}
\left(\pp{t}{h}\right)_{\mathcal{S},\alpha}
\quad\text{and}\quad
\left(\pp{t}{\alpha}\right)_{\mathcal{S},h}
\ ,
\label{rates}
\end{align}
where the first expression quantifies the change of the temperature 
$t$ with field and the second expression quantifies 
the change of the temperature $t$ with pressure
on a two-dimensional isentrope surface ($\mathcal{S}$ isosurface).

\begin{figure}[h]
\centering
\includegraphics*[width=0.85\columnwidth]{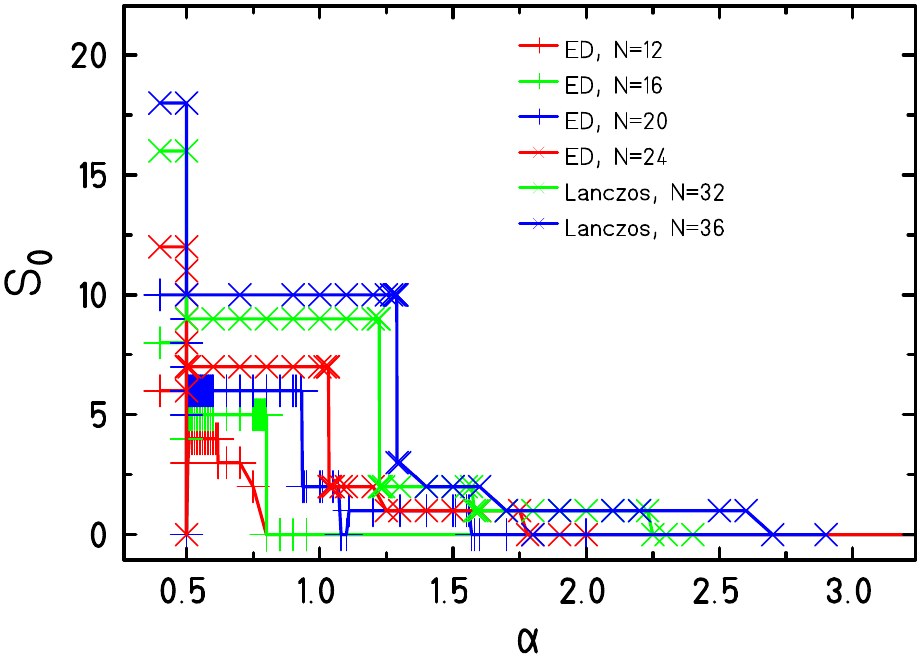}

\includegraphics*[width=0.85\columnwidth]{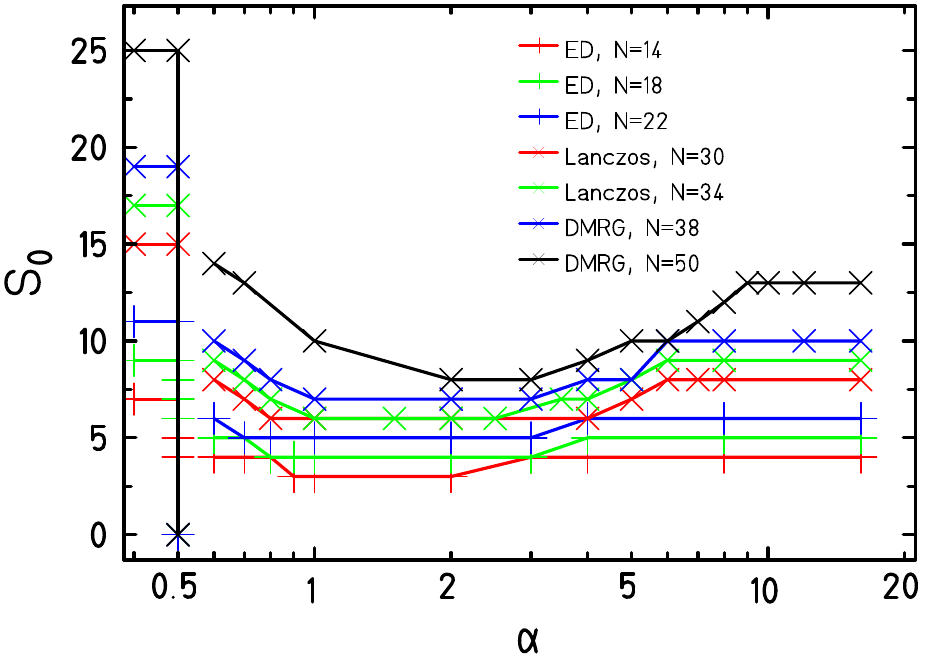}
\caption{Ground-state spin quantum number $S_0$ for various $\alpha$ and selected 
lengths $N$ that are multiples of four (top) or multiples of two, but not of four
(bottom). Depending on the length exact diagonalization (ED), 
Lanczos diagonalization (Lanczos),
or SU(2) density-matrix renormalization group theory (DMRG) was employed.
The lines are drawn as a guide for the eye.
}
\label{ground-state-spin}
\end{figure}

\section{Results}
\label{sec-3}

\subsection{Ground state properties and quantum critical point}
\label{sec-3-A}

For $0\leq \alpha= J_2/|J_1| < \alpha_c = s_a/(2 s_b)$ the ground state is given
by the multiplet of the fully polarized, i.e.\ ferromagnetic state. 
For $N$ spins of spin quantum number $s=1/2$ this would yield a ground state spin
of $S_0=N/2$.
For $\alpha > \alpha_c$ the character of the ground state depends on whether the
lengths of the sawtooth chain (with periodic boundary conditions) is a
multiple of four or a multiple of two but not of four, 
at least for $s=1/2$,
compare \cite{ToK:JMMM04,RPP:SP23}. 
Figure~\xref{ground-state-spin} displays this behavior for a selection of smaller
systems (always with periodic boundary conditions). The result of the top panel was 
already discussed in Ref.~\cite{RPP:SP23}: The ground state spin drops to zero 
for large $\alpha$ in case of sawtooth chains where the length is a multiple of four.
Cases of even length where the length is not divisible by four behave differently.
Here the ground state spin assumes a non-zero value for large $\alpha$; 
for $N$ spins $s=1/2$ this would be $(N+2)/4$. The underlying reason is that the
basal spins form a spin ring of odd length whose approximate ground state spin
is $1/2$ for large $\alpha$ \cite{BSS:JMMM00:B}. 
The apical spins are then polarized accordingly due to the ferromagnetic 
interaction.

\begin{figure}[h]
\centering
\includegraphics*[width=0.85\columnwidth]{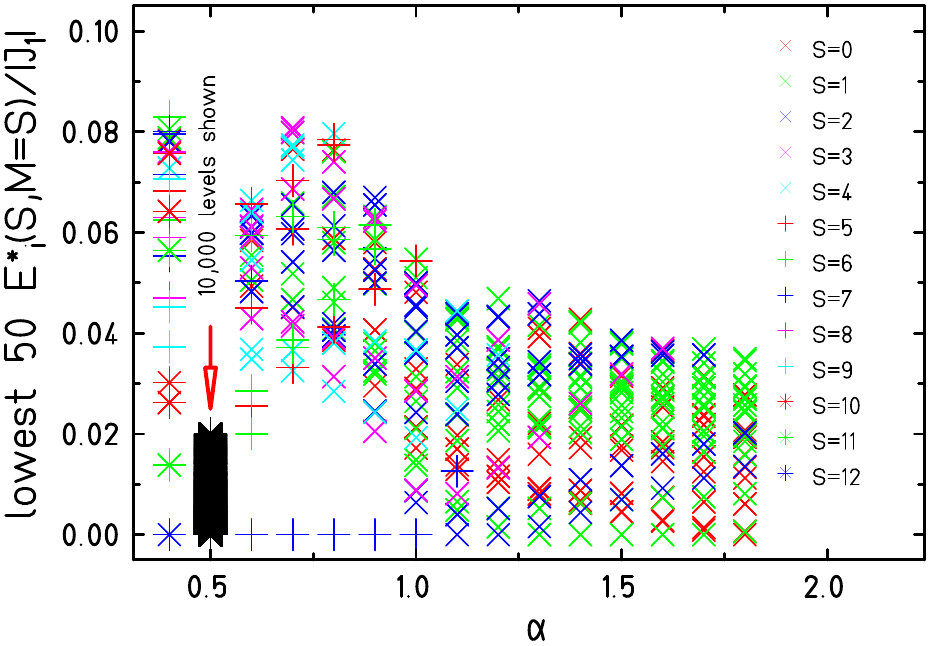}
\caption{Example of low-lying energy levels for $N=24$ as function of $\alpha$:
Only the lowest 50 eigenvalues are shown for selected values of $\alpha$
except for $\alpha=1/2$, where the lowest 
10,000 eigenvalues are displayed. The intention is to picture the 
very untypical density of states at the QCP for $\alpha=1/2$.
}
\label{dos-24}
\end{figure}

For $\alpha = \alpha_c$ the system features a quantum critical point (QCP)
with macroscopic ground state degeneracy \cite{KDN:PRB14,DSD:EPJB20}
and a very dense low-lying spectrum.
Figure~\xref{dos-24} displays this behavior for the case of 24 spins $s=1/2$.
The lowest 50 eigenvalues are shown for selected values of $\alpha$ 
except for $\alpha=1/2$, where the lowest 10,000 eigenvalues are displayed.
The collapse of the spectrum for $\alpha\rightarrow\alpha_c$ is the reason for 
the remarkable caloric properties of the system in the vicinity of this 
quantum critical point.
The degenerate ground state contains the multiplets of many spin quantum 
numbers, $S_0 \in \{0, S_{0,\text{min}}, \dots, S_{0,\text{max}}\}$, where e.g.\
for (all even) $N$ spins $s=1/2$ $S_{0,\text{max}}=N/2$ and for $N$ being 
a multiple of four $S_{0,\text{min}}=N/4+1$.

\begin{figure}[ht!]
\centering
\includegraphics*[width=0.85\columnwidth]{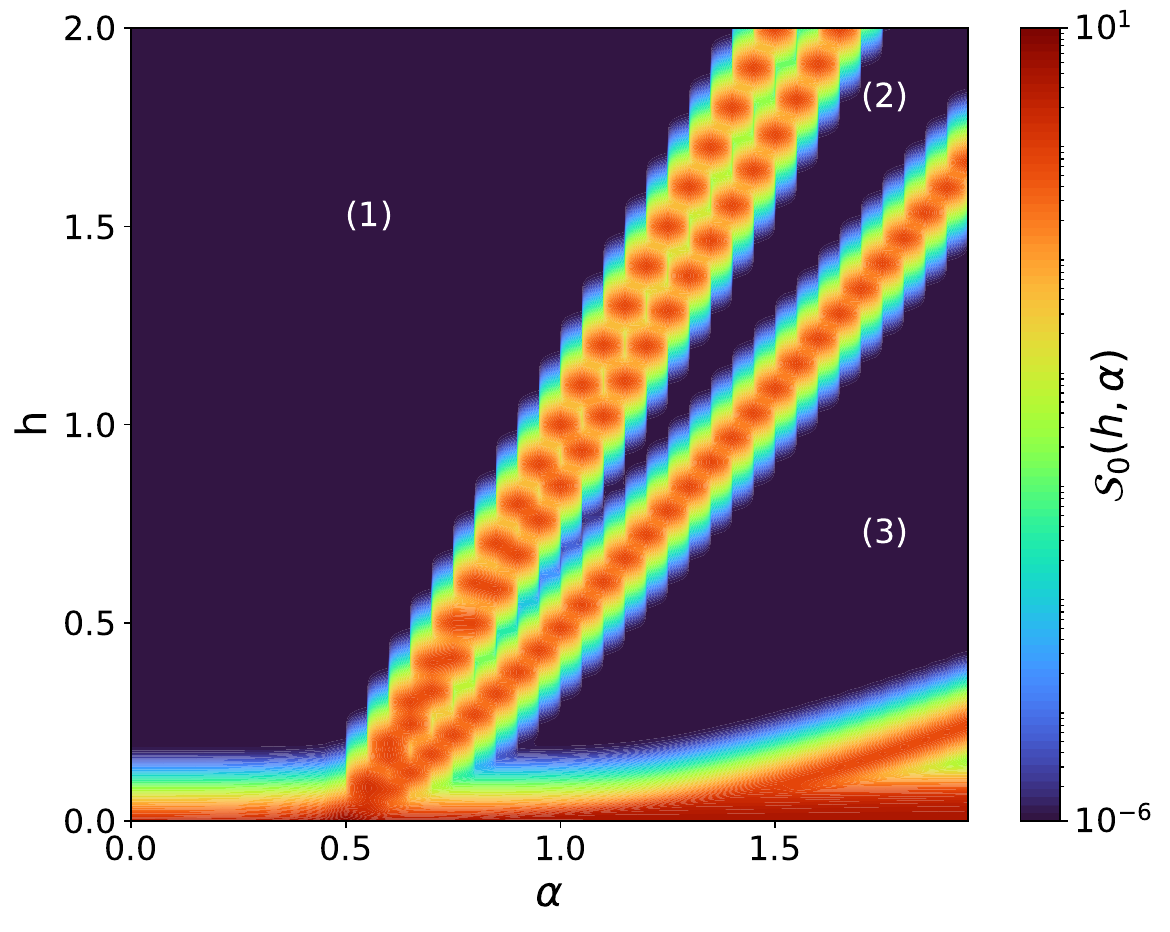}
\caption{Map of (near) ground state entropy for $N=16$ and $s=1/2$ 
as function of $\alpha$ and $h$ (logarithmic scale).
The entropy was evaluated at $t=0.01$ for technical reasons.
At $\alpha=0.5, h=0$ the quantum critical point is visible. 
The dotted curves that move towards the upper right corner
separate areas with ground state spins 
$S_0 \in \{S_{0,\text{max}}, \dots S_{0,\text{min}}\}$ from left to right; 
here in particular $S_0=8$ in (1), $S_0=6$ in (2), and $S_0=5$ in (3), 
compare also \figref{gs-16-1}.
The light region for larger $\alpha>1$ and 
almost vanishing field $h$ reflects a growing low-lying density of states
consisting of excitations with $S \in \{0, \dots 4\}$.}
\label{entropy-16-1}
\end{figure}

\subsection{Magneto- and barocaloric phase diagram}

\begin{figure}[ht!]
\centering
\includegraphics*[width=0.85\columnwidth]{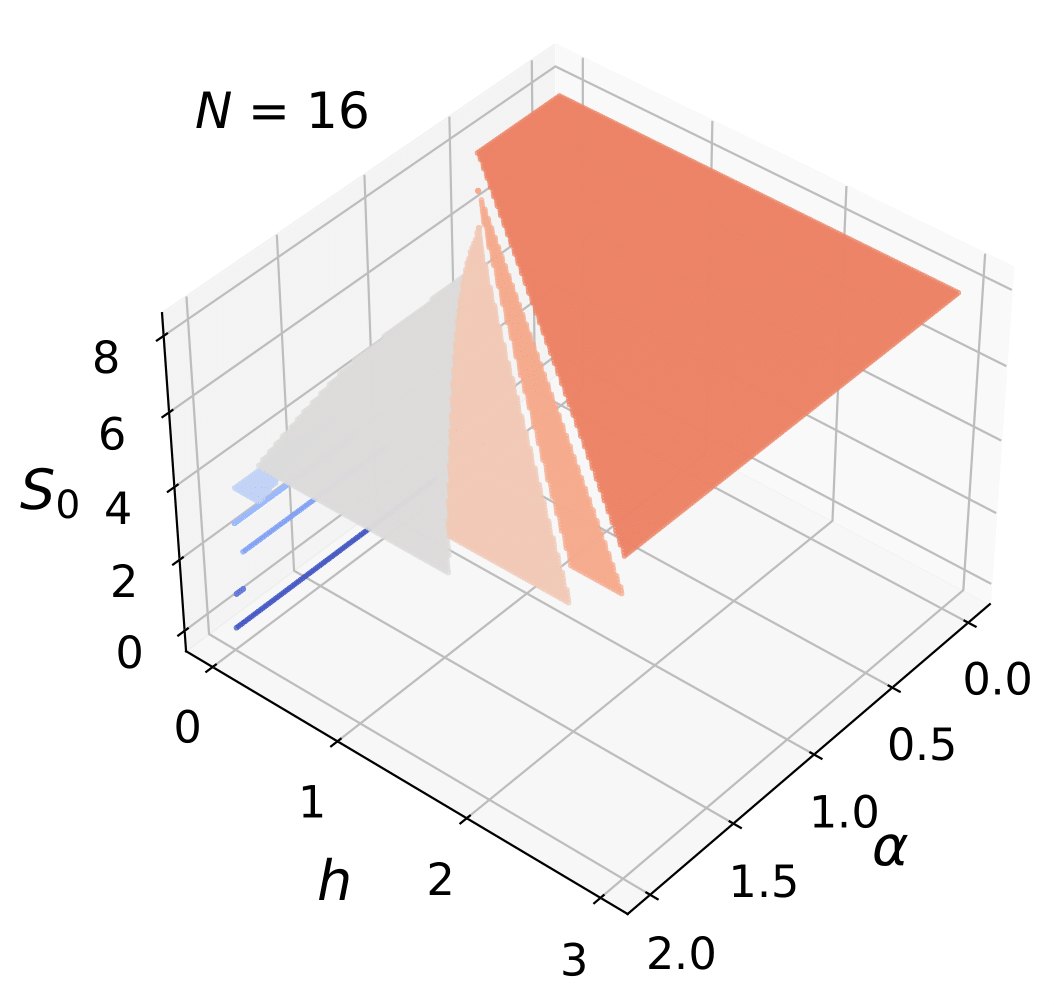}
\caption{Ground state spin $S_0$ for $N=16$ and $s=1/2$ as function of 
$\alpha$ and $h$.
}
\label{gs-16-1}
\end{figure}

While the quantum phase transition at $\alpha = \alpha_c, h=0$ 
yields a macroscopic ground state degeneracy, 
further points of ground state degeneracy can be found in the $\alpha$-$h$--plane. 
For each $\alpha$, an external magnetic field $h$ splits 
all multiplets according to the respective 
magnetic quantum number $M$. 
If $M=-N/2$ (for $N$ spins with $s=1/2$) is not already the energetically
lowest state for $\alpha>0.5$, 
compare example in \figref{dos-24},
crossings appear whenever magnetic levels of higher-lying
multiplets become new ground states. For those multiplets that are
part of the degenerate ground state at the QCP, this yields curves 
of level crossings
in the $\alpha$-$h$-plane that head towards the QCP,
compare \figref{entropy-16-1} and \figref{gs-16-1}
as well as discussion below.

The whole scenario can be pictured by means of an entropy map as displayed 
in \figref{entropy-16-1} for the case of $N=16$ and $s=1/2$. 
The near ground state entropy is color-coded on a logarithmic scale; 
it was evaluated at $t=0.01$ for technical reasons.
At $\alpha=0.5, h=0$ the quantum critical point is visible. 
The reddish curves separate areas with ground state spins 
$S_0 \in \{S_{0,\text{max}}, \dots S_{0,\text{min}}\}$ from left to right. 
For larger $\alpha>1$ and almost vanishing field $h$ 
the reddish curves reflect a with $\alpha$ 
growing low-lying density of states
consisting of excitations with smaller total spin, here 
$S \in \{0, \dots 4\}$. 
In addition, the spectrum splits into bunches of levels 
with further increasing $\alpha$ since the system progressively turns into 
an antiferromagnetic ring and loosely attached, almost free apical spins,
compare also \cite{KrD:JPCM22}.
This is reflected by the upturning branch at low fields 
and $\alpha > 1.25$.

\begin{figure}[ht!]
\begin{center}
\includegraphics[clip=on,width=70mm,angle=0]{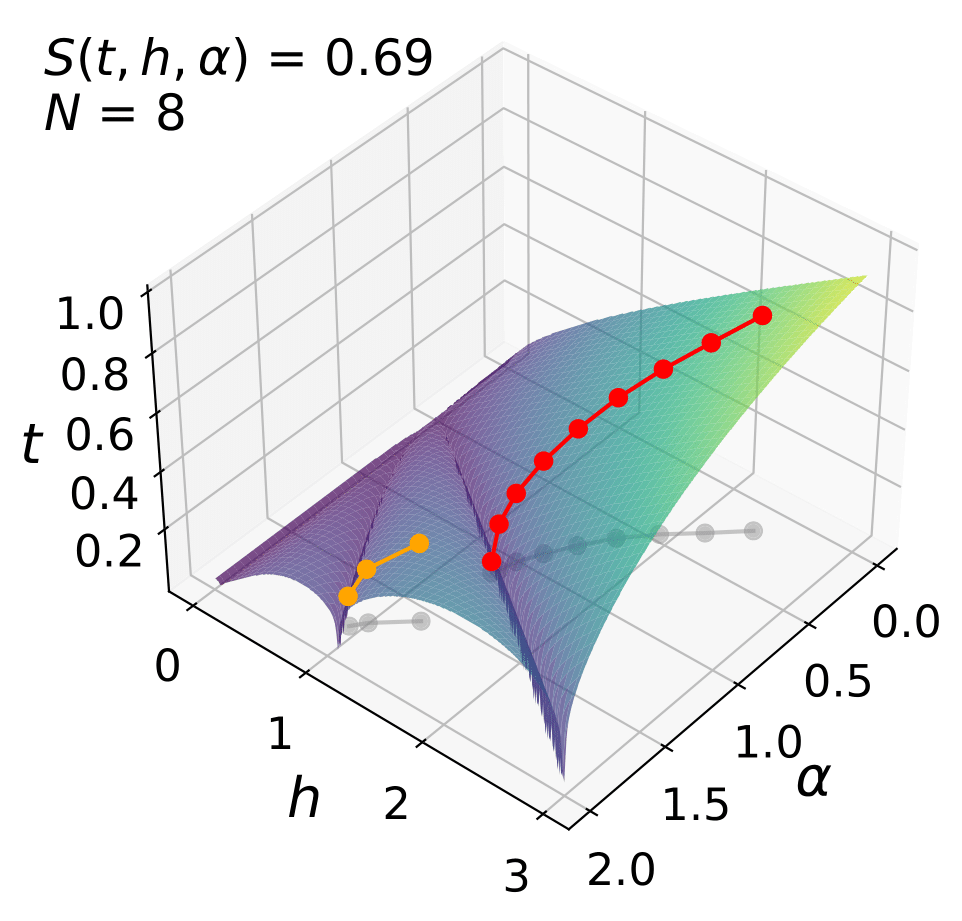}
\caption
{Isentrope surface with $S(t, h, \alpha) = 0.69< \log(2)\approx 0.693$ 
in the parameter-space of $\alpha, h$, and $t$. 
The dotted curves represent processes of steepest descent 
on which cooling rates assume maximal values.}
\label{F3} 
\end{center}
\end{figure}

\begin{figure}[ht!]
\begin{center}
\includegraphics[clip=on,width=60mm,angle=0]{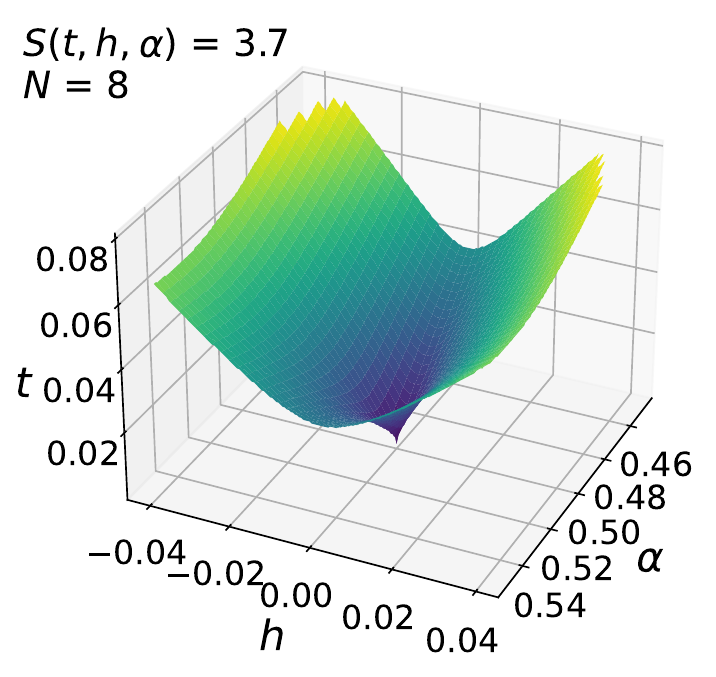}

\includegraphics[clip=on,width=60mm,angle=0]{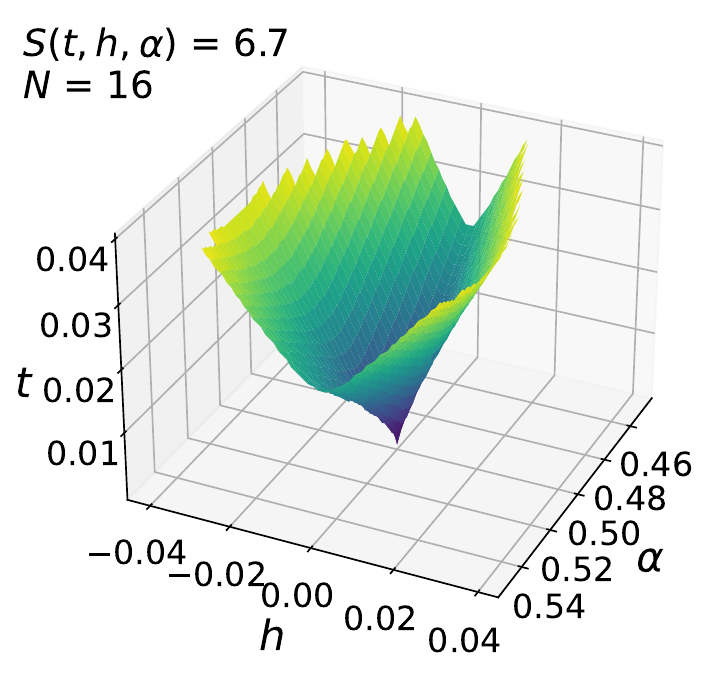}

\includegraphics[clip=on,width=60mm,angle=0]{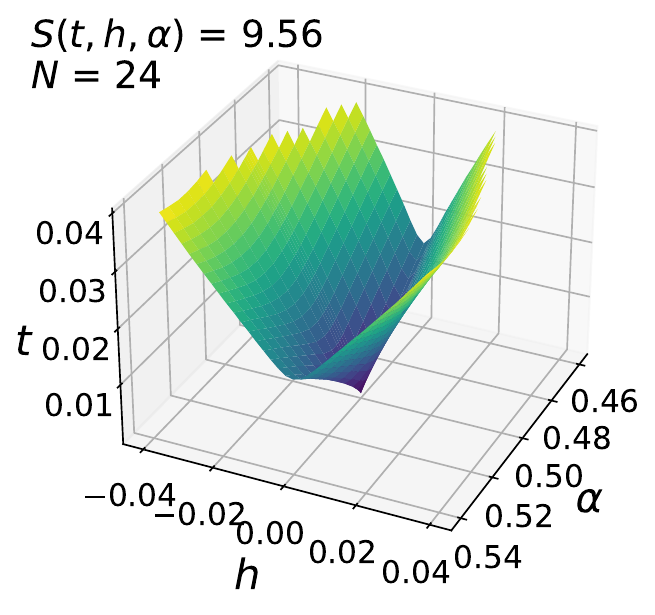}
\caption
{Isentrope surfaces in the parameter-space of $\alpha, h$, and $t$ 
in the vicinity of the QCP
for $N=8$ (top), $N=16$ (middle), and $N=24$ (bottom) and entropy values
close to the residual entropy at the QCP. $N=8$: $S=3.7 < \log(41)\approx 3.714$,
$N=16$: $S = 6.7 < \log(817)\approx 6.706$,
$N=24$: $S=9.56<\log(14260)\approx 9.565$.}
\label{F7} 
\end{center}
\end{figure}

The change of ground state spin quantum numbers can also be assessed 
by looking at \figref{gs-16-1}, where $S_0$ is displayed in terms of
colored planes as function of $\alpha$ and $h$. 
The above discussed ground state degeneracies
happen at the steps between planes of constant $S_0$ of which the rightmost
three run towards the QCP at $\alpha=0.5, h=0$ since the respective multiplets
are part of the degenerate ground state.

Moving from zero to non-zero temperatures the systems allows many interesting
thermodynamic processes in particular since the iso-entropy (isentropic) surfaces
are strongly curved and feature cusps for low temperatures, compare \figref{F3}
for a small system with $N=8$ and $s=1/2$. In contrast to purely barocaloric or
purely magnetocaloric systems, a magnetic quantum critical system offers 
processes with very large cooling rates when pressure and field are modified 
simultaneously in order to achieve a steepest descent on an isotrope surface. 
Two such processes are depicted in \figref{F3} by curves of connected 
bullet points.

A close-up look at the vicinity of the quantum critical point in \figref{F7} 
shows that isentropes close to the value corresponding to the degeneracy
of the QCP assume very large slopes towards zero temperature. These isentrope surfaces mark
the quantum critical region inside which one observes universal behavior, 
i.e.\ no scale is provided by system parameters, only temperature controls the behavior,
and the fluctuations possess classical character, see also
\cite[Fig. 1 (r.h.s.)]{Voj:AdP00} 
or \cite[Fig. 1 (l.h.s.)]{Voj:RPP03}. 
For $\alpha\ne 0.5$, a system-specific low-energy scale emerges due to frustration
that can be orders
of magnitude smaller than the involved parameters of the system such as $J_{1/2}$
\cite{OLI:SSC88,Ram:ARMS94}. Here for $\alpha > 0.5$, quantum fluctuations dominate 
whereas for $\alpha < 0.5$  a ferromagnetic ordering exists at zero temperature 
only \cite{Voj:AdP00}. As one can see in \figref{F7} and in accordance with \cite{Voj:AdP00}
the quantum critical region gets narrower with increasing system size. 
This is related to the fact that the residual entropy for $\alpha\ne 0.5$ scales 
roughly like $\log(N)$ whereas for $\alpha = 0.5$ at the QCP it scales 
approximately like $N$.
This suggests that the largest cooling rates are only achievable in very close 
proximity to the QCP.

\section{Discussion}
\label{sec-4}

Materials that are susceptible to pressure and external magnetic fields 
allow the combined use of both for caloric processes, compare
Refs.~\cite{GSZ:PRL20,JCF:N21,WLX:PRL23} for related recent examples.
This might, e.g.,\ be advantageous 
when running single-shot cooling experiments or thermodynamic cycles with field and pressure sweeps. 
Spin systems such as the
discussed ferromagnetic-antiferromagnetic sawtooth chain offer the additional 
feature of a quantum critical point, whereby in the vicinity of this point 
cooling rates are particularly large. In the space of external parameters 
$\alpha$ (related to pressure) and $h$ (magnetic field) this special
point can be approached from various directions, not only via change of
pressure which provides more experimental options.

Quite recently, the idea to employ magneto-electric couplings was formulated
for the sawtooth chain in order to drive the ratio of the two exchange interactions 
towards the critical value \cite{ROS:PRB22}. This would enable one to 
use the material more closely to the QCP and thus to benefit from 
much larger cooling rates. Altogether, the electric field 
could replace pressure to drive $\alpha$.
One could then speak of a magneto-electric drive where one can apply 
both magnetic as well as electric fields to steer the thermodynamic
processes.

\section*{Acknowledgment}

This work was supported by the Deutsche Forschungsgemeinschaft DFG
(355031190 (FOR~2692); 397300368 (SCHN~615/25-2);  
RI~615/25-1 and 449703145 (SCHN~615/28-1)).


%

%
\end{document}